# Quantum Information Scrambling in Molecules

Chenghao Zhang,[1] Peter G. Wolynes[2,*] and Martin Gruebele[1,3,*]

[1]Department of Physics, University of Illinois at Urbana-Champaign, Illinois 61801, USA,
[2]Departments of Chemistry and Physics, and Center for Theoretical Biological Physics, Rice University, 6100 Main Street, Houston, TX 77251, USA, [3]Department of Chemistry, Center for Biophysics and Quantitative Biology, University of Illinois at Urbana-Champaign, Illinois 61801, USA

*Corresponding authors: pwolynes@rice.edu, mgruebel@illinois.edu

*Out-of-time-order correlators (OTOCs) can be used to probe how quickly a quantum system scrambles information when the initial conditions of the dynamics are changed. In sufficiently large quantum systems, one can extract from the OTOC the quantum analog of the Lyapunov coefficient that describes the time scale on which a classical chaotic system becomes scrambled. OTOCs have been applied only to a very limited number of toy models, such as the SYK model connected with black hole information scrambling, but they could find much wider applicability for information scrambling in quantum systems that allow comparison with experiments. The vibrations of polyatomic molecules are known to undergo a transition from regular dynamics at low energy to facile energy flow at sufficiently high energy. Molecules therefore represent ideal quantum systems to study scrambling in many-body systems of moderate size (here 6 to 36 degrees of freedom). By computing quantum OTOCs and their classical counterparts we quantify how information becomes 'scrambled' quantum mechanically in molecular systems. Between early 'ballistic' dynamics, and late 'saturation' of the OTOC when the full density of states is explored, there is indeed a regime where a quantum Lyapunov coefficient can be defined for all molecules in this study. Comparison with experimental rate data shows that slow scrambling as measured by the OTOC can reach the time scale of molecular reaction dynamics. Even for the smallest molecules we discuss, the Maldacena bound remains satisfied by regularized OTOCs, but not by unregularized OTOCs, highlighting that the former are more useful for discussing information scrambling in this type of moderate-size quantum system.*

## I. INTRODUCTION

At low excitation energy, molecules are described by good quantum numbers. In contrast, at energies sufficient for chemical reactions, it has long been thought that the dynamics of molecules can be treated statistically. Already in 1919, on the basis of the old quantum theory [1], Herzfeld argued that the maximum rate of rearranging atoms in a thermalized molecule was $k_B T/h$. A related bound for the rate of scrambling quantum information named after Maldacena [2] has recently emerged from the study of black holes [3], string theory [4] and many body localization [5].

Determining the rate of quantum information scrambling has been made precise by using out of time order correlation functions (OTOCs) [6]. Quantum OTOCs can be constructed as analogs of the classical Lyapunov exponents that measure the instability of classical trajectories arising from perturbations in the initial conditions. In this paper, we use OTOCs to quantify how rapidly quantum information is scrambled in molecules. Molecular vibrations are attractive for building local random matrix models of quantum scrambling [7] because of their detailed connection with accurate experiments [8,9], relative to more abstract models such as the Sachdev-Ye-Kitaev (SYK) model [10], which currently lack a direct connection with experiment. In addition, the dynamics of molecules can be tuned from a regime described by well-defined quantum numbers to the statistical regime by varying the energy content or the size of the molecules being studied [7,11–16]. The key question then becomes: could slow or incomplete scrambling, as measured by OTOCs, be slower than the barrier-crossing or photodissociation time, and thus interefere with statistical behavior in an atom-rearranging molecular reaction? There are certainly experimental examples where the inability to scramble quantum numbers sufficiently seems to limit molecular reaction rates [17–21]. OTOCs then become useful tools to determine when molecules can be treated statistically. Here we compute OTOCs for several molecules where quantum scrambling occurs through Fermi resonances, higher order anharmonic resonances. or Coriolis coupling [22–29]. We compare the quantum dynamics with the classical limit using Lyapunov stability analysis [30,31], make a connection between scrambling rates and experimental bounds on reaction rates of molecules, and comment on how the vibrational dynamics of molecules obeys the Maldacena bound depending on the type of OTOC being used to assess quantum scrambling.

## II. METHODS
### A. Model Hamiltonian

Optical excitation of molecules deposits energy into "bright states" with specific quantum numbers. The subsequent spreading of the wave function then populates the state space [7,25]. Many of the $f \equiv 3N - 6$ molecular vibrations of a typical $N$-atom organic molecule are still in the quantum limit at room temperature. Owing to the Born-Oppenheimer approximation, the resulting potential surface is smooth, so that these approximately harmonic vibrations are coupled by individually weak nonlinearities whose strength decreases with the order $m$ of coupling, scaling approximately as $\phi_m = (-1)^m \phi_3 \gamma^{m-3}$ ($m \geq 3$, $\gamma \approx 0.1$ to $0.3$ ) [24] (see Appendix A for details). The Fermi resonant term $m$=3 was first noticed in carbon dioxide [32]. Usually, the magnitude of $\phi_3$ lies in the weak coupling limit ($\phi_3 \leq 0.1\omega_i$). Yet, molecular rate theories generally assume that thermalization is very fast because most molecules have many modes, such that statistical models can be applied.

The molecular vibrational Hamiltonian describes the interchange of energy among the *f* modes of a molecule. It can be expressed using a Fock space representation in terms of the occupancy of these modes as $H = H_0 + V$, where

$$H_0 = \sum_{i=1}^{f} \epsilon_i(\hat{n}_i) \,, \quad V = \sum_{\boldsymbol{m}} \prod_i \phi_{\boldsymbol{m}} a_i^{\dagger \, m_i^+} a_i^{m_i^-}. \tag{1}$$

$H_0$ describes the uncoupled motion of a set of oscillators with mode energies $\epsilon_i$, while $V$ describes the anharmonic couplings between them. The unperturbed modes may be allowed to be anharmonic themselves through $\epsilon_i(\hat{n}_i)$. $a_i$ and $a_i^\dagger$ are the ladder operators for mode *i*. The index $\boldsymbol{m}$ gives quantum number differences $m_i^+$ and $m_i^-$ between anharmonically coupled states, which add up to the order *m* of the coupling (see Appendix A for details). For one of our molecular models, we also include the effect of Coriolis couplings, which could lead to additional scrambling due to vibration-rotation interactions on time scales longer than the vibrational scrambling of interest here [33–35].

## B. OTOCs and Lyapunov coefficients

We compute the whole quantum Lyapunov spectrum, which reflects the Kolmogorov-Sinai entropy of the dynamics [36] and its classical counterpart. From these one can extract the largest exponent at early time. Suppose phase space is described by coordinates $\boldsymbol{x}$ and their conjugate momenta $\boldsymbol{p}$, where $z_i (i = 1, \cdots, 2f)$ denotes *x* or *p* in general. In classical mechanics, the sensitivity to initial conditions is captured by the Lyapunov spectral matrix $L_{ij}^{cl}$ given by the Poisson bracket matrix $M_{ij}^{cl}$ as

$$L_{ij}^{cl}(t) = \left((M^{cl})^\dagger(t) M^{cl}(t)\right)_{ij} \text{ with } M_{ij}^{cl}(t) \equiv \frac{\partial z_i(t)}{\partial z_j(0)}. \tag{2}$$

The quantum analogs can then be defined as an out-of-time order correlation

$$\hat{L}_{ij}(t) = \left(\widehat{M}^\dagger(t)\widehat{M}(t)\right)_{ij} \text{ with } \widehat{M}_{ij}(t) \equiv [\hat{z}_i(t), \hat{z}_j(0)], \tag{3}$$

where the classical Poisson brackets have become commutators and a variety of averages can be employed. We find that reformulating $\widehat{M}_{ij}$ in terms of ladder operators associated with $z_i$ yields the best concordance for our systems between the quantum and classical pictures (see Appendix B for details). We diagonalize both $\hat{L}_{ij}(t)$ and $L_{ij}^{cl}(t)$ to obtain time-dependent Lyapunov eigenvalues $s_i(t)(i =$ $1,\cdots,2f$) (or $s_i^{cl}(t)$). This representation of the OTOC is convenient because the Heisenberg relation provides the normalization $s_i(0) = 1$, so growth can be monitored relative to the initial condition. For $s_i(t)$ we consider three time scales: The first time scale, $\tau_b$, is the "ballistic time" required to initiate scrambling classically, or quantum mechanically for the survival probability $P(t)$ to go outside its initial quadratic phase $P \sim 1 - at^2$ which arises from the coupling of discrete levels [37]. The second time scale, $\tau_\lambda = \lambda_1^{-1}$, is the inverse of the largest Lyapunov coefficient, the rate at which the initially encoded information of the system gets scrambled. The last time scale, $\tau_s$, is the 'scrambling time', at which the largest eigenvalue $s_1(t)$ levels off and the quantum system has reached maximum scrambling. (Three time scales are indicated in Figure 1 for a specific model.) The Lyapunov exponents $\lambda_i(t)$ (and by analogy $\lambda_i^{cl}(t)$) can then be defined as $\lambda_i(t) = \frac{1}{2}\partial ln(s_i(t))/\partial t$ in the region $\tau_b < t < \tau_s$, where we evaluate the derivative using a smoothing spline fit to $s_i(t)$.

To compute $\hat{L}_{ij}(t)$ for a given initial state, we numerically solve the time-dependent Schrödinger equation using the shifted-update-rotation (SUR) algorithm [38]. To compute the classical Lyapunov spectrum, we use the Bulirsch-Stoer algorithm [39] to integrate Hamilton's equations of motion. For each molecular system, we calculate classical trajectories starting from 100 randomly chosen angles ($\theta_i$), but with action value ($J_i$) that correspond to the quantum numbers $\{n\}$ of the initial state used to compute the quantum OTOC (see Appendix C for computational details and convergence checks).

### III. RESULTS
#### A. OTOCs for 4 molecular systems

We first study the Schofield-Wyatt-Wolynes (SWW)-Hamiltonian [40]. Its *f*=6 anharmonically coupled Morse oscillators stylistically represent the pairwise-coupled local C-C stretches in a benzene ring [41,42] (see Appendix D for details). To simplify computation of *z*(*t*), here we set the self-anharmonicity to 0 and work with anharmonically coupled harmonic oscillators. For illustration we choose the state with quantum numbers (2,2,3,2,2,2) at 13,000 cm$^{-1}$ as the initial state and $\phi_3 = 7$ cm$^{-1}$. (We use *E/hc* units throughout, conventional in vibrational spectroscopy.) While the model is in the weak coupling limit, the near-degeneracy of modes facilitates efficient scrambling. Figure 1 compares the largest eigenvalues of the *L* matrices and the corresponding Lyapunov exponents (shown as an inset) for the quantum and classical cases. At early times (0.3 ps < *t* < 1.5 ps) the slopes of $s_1^{cl}(t)$ and $s_1(t)$ correspond closely. Thus, we observe a clear quantum-classical correspondence for this model in the weak coupling regime.

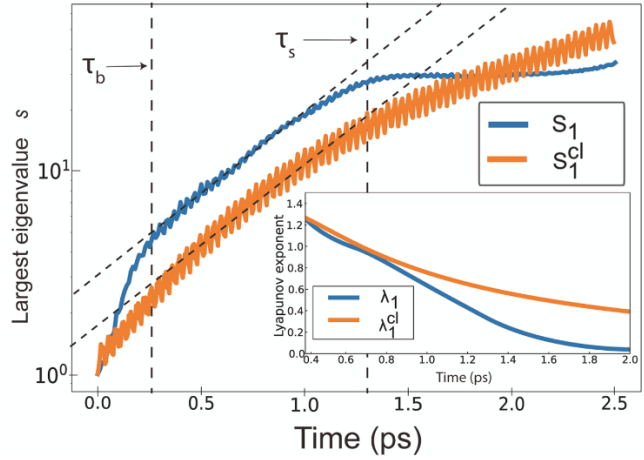

**Figure 1.** Color online. The largest eigenvalue $s$ of $L_{ij}^{cl}(t)$ (orange) and $\hat{L}_{ij}(t)$ (blue) for the SWW model. Initial state: (2,2,3,2,2,2), anharmonic coupling $V = 7$cm$^{-1}$. The inverse of the largest classical Lyapunov coefficient extracted from the plot is $t_0 = 1$ ps. We see $s_1(t)$, $s_1^{cl}(t)$ agree well up to 1.3 $t_0$. Inset: Largest classical (orange) and largest quantum (blue) Lyapunov exponent extracted from $s_1^{cl}(t)$ and $s_1(t)$. The classical Lyapunov coefficient levels off more slowly than the quantum one. Note that the decay of $\lambda_1^{cl}(t)$ to a smaller asymptote is not due to a finite $\Delta z_i(t=0)$, but due to exploration of a less chaotic region as time grows. This is also true in Figures 2a and 3. The ballistic regime ($\tau_b$) merges into the Lyapunov regime (slope $\lambda_1 = \tau_\lambda^{-1}$) at the leftmost dashed line. The Lyapunov regime merges into the scrambling regime ($\tau_s$) at the rightmost dashed line.

We next study an $f=8$ model with strong stretch-bend resonances that describes the CH stretches and HCH bending overtones of the molecule cyclopentanone in the 2900 cm$^{-1}$ region. This model manifests a strong resonance with $\phi_3 = 25\ cm^{-1}$, and includes up to 4$^{th}$ order couplings (see Appendix D for details). Such vibrational resonances play an important role for onset of chaos in the Arnold web [43,44]. We study the OTOC for two initial states, (1,1,1,0,0,1,0,0) at ~9500 $cm^{-1}$ where the system is weakly coupled, and (2,2,3,2,2,2,2,2) at ~43700 $cm^{-1}$ where the system is strongly coupled. For the weakly chaotic low energy case shown in Figure 2(a), there is good agreement between Lyapunov exponents $\lambda_1^{cl}(t)$, $\lambda_1(t)$, as was the case for the SWW Hamiltonian. For the second example, $s_1^{cl}(t)$ very quickly deviates from $s_1(t)$, even before $s_1(t)$ reaches its plateau: at this higher energy, most invariant tori have been destroyed and classical trajectories very quickly enter the chaotic region of phase space, whereas the quantum system undergoes quantum localization [11] that prevents growth of $s_1(t)$: as energy spreads throughout the modes, the energy per mode of the high frequency modes is reduced, resulting in weaker coupling of the effectively more harmonic modes [13]. A similar instability of classical motion compared to its quantum counterpart was discussed in [45].

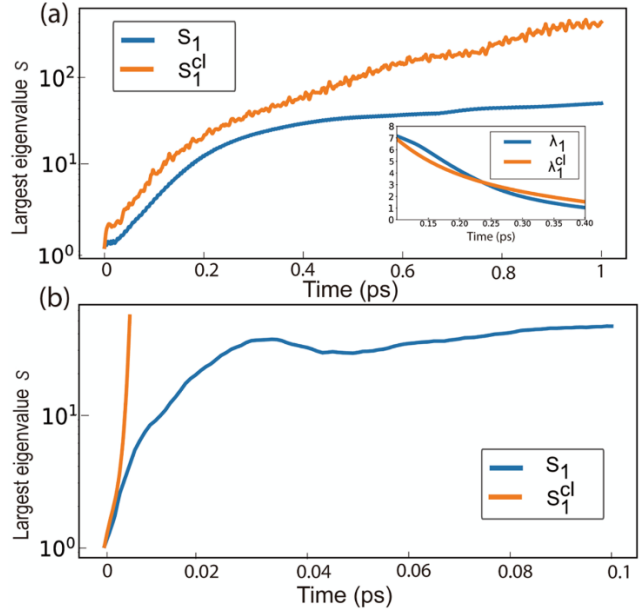

**Figure 2.** Color online. Hamiltonian taken from [22]. **(a)** Main figure: $s_1^{cl}(t)$ (orange) and $s_1(t)$ (Blue) for the 8-mode cyclopentanone based model in the weakly chaotic (low energy) region. $T_0 \equiv 1/\lambda = 0.2$ ps. Initial state: (1,1,1,0,0,1,0,0). Similar to Fig.1, we see agreement for $s_0^{cl}(t)$ and $s_0(t)$ before quantum OTOC reach its plateau at $t = 2T_0$. Inset: $\lambda_0^{cl}(t)$ (orange) and $\lambda_0(t)$ (blue). We see $\lambda_0(t)$ and $\lambda_0^{cl}(t)$ are in good agreement at $t < 2T_0$. **(b)** $s_1^{cl}(t)$ (orange) and $s_1(t)$ (Blue) for the cyclopentanone-based model in the strongly chaotic (high energy) region. $T_0 = 0.01$ ps. Initial state: (2,2,3,2,2,2,2,2). We see $s_1^{cl}(t)$ grows rapidly and deviates from $s_1(t)$ before $s_1(t)$ reaches a plateau not very different from the lower energy state.

We also computed the Lyapunov spectrum for two full-dimensional models of vibrating molecules, cyclopentene with all of its 33 modes [46] and cyclopentanone with its 36 modes [47] (see Appendix D for details). These begin to approach in size SYK model numerical simulations. In an analysis of quantum beat experiments, Bigwood *et al.* [22] have shown that cyclopentene and cyclopentanone display an onset of facile energy flow near the energy of the C-H overtone. At our chosen energy of ~17500 cm$^{-1}$, above this overtone energy, we verified that the initial states satisfy the energy flow criterion deduced by Logan and Wolynes [7], and that $P(t)=|\langle 0|t \rangle|^2$ rapidly decays to a small inverse participation number or 'dilution factor' $N_p^{-1}=\sigma$ (see Table 1 in Appendix for this and all other model systems). In their theory, the transition to facile energy flow is not determined by the total density of states $\rho_{tot}$, but rather by the criterion $\rho_{loc} V_{anh} > 1$, where $V_{anh}$ is the local anharmonic coupling strength, and $\rho_{loc}$ is the 'local' density of states that are directly coupled by the $V_{anh}$ terms in the Hamiltonian; when the criterion is satisfied, energy flows freely because the spacing of locally coupled energy levels becomes comparable to the local anharmonic coupling.

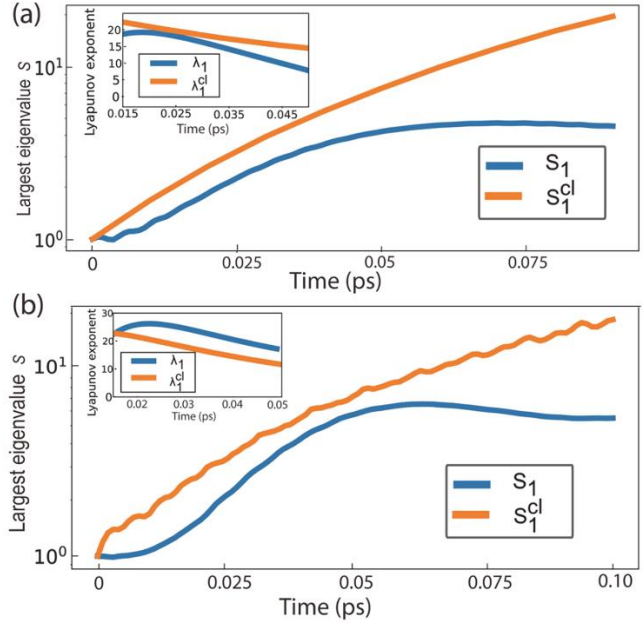

**Figure 3.** Color online. (a). Main figure: $s_1^{cl}(t)$ (orange) and $s_1(t)$ (Blue) for cyclopentene at $E = 17569 cm^{-1}$. Inset: $\lambda_0^{cl}(t)$ (orange) and $\lambda_0(t)$ (blue). $T_0 \equiv 1/\lambda = 0.05 ps$. $\lambda(t)$ is in unit of $ps^{-1}$. (b) Main figure: $s_1^{cl}(t)$ (orange) and $s_1(t)$ (Blue) for cyclopentanone at $E = 17357 cm^{-1}$. **Inset:** $\lambda_0^{cl}(t)$ (orange) and $\lambda_0(t)$ (blue). $T_0 \equiv 1/\lambda = 0.05$ ps. We see in these two large organic molecules, quantum and classical Lyapunov exponents are close to each other until $t = T_0$.

Like the SWW model, simulation of scrambling in these organic molecules shows good agreement between $L^{cl}(t)$ and $\hat{L}(t)$ for $\tau_b < t < \tau_s$ in Fig. 3, but the OTOCs for the quantum systems level off at the scrambling time ($\tau_s \approx 0.05$ ps).

In near-integrable systems, the KAM theorem dictates that the motion for the majority of initial conditions (the 'regular set') will be along invariant tori [48–50]. Chaotic islands are rare and lie at resonance junctions in phase space. This phenomenon has been observed in simulations of the Arnold web for several molecular Hamiltonians [26,44,51–53], and we illustrate this case with the small organic molecule $SCCl_2$, whose Arnold web has been studied [43,54]. The $f=6$ mode Hamiltonian for $SCCl_2$ is taken from [35]. In contrast to what was seen for the large molecules cyclopentene and cyclopentanone, the classical simulations of $SCCl_2$ are regular in most regions, while the quantum Lyapunov spectrum shows early growth. Classical simulations for $SCCl_2$ starting initially with actions $(S_i/h)=(6, 5, 1, 3, 5, 3)$, along with 100 randomly chosen initial angles, were used to find the average and largest values of $s_1^{cl}(t)$, which are compared with their quantum counterpart $s_1(t)$ in Figure 4(a) (see Appendix C,E for details). The OTOCs $\hat{L}_{ij}(t)$ and $L_{ij}^{cl}(t)$ for $SCCl_2$ behave quite differently: The growth of the classical stability eigenvalue is delayed initially, until the molecule escapes from the regular region, and then the OTOC grows very rapidly. The quantum wave packet samples more state space initially, so that $\hat{L}_{ij}(t)$ rapidly inflates at early times but then levels off at long times, when the quantum system exhausts the accessible states.

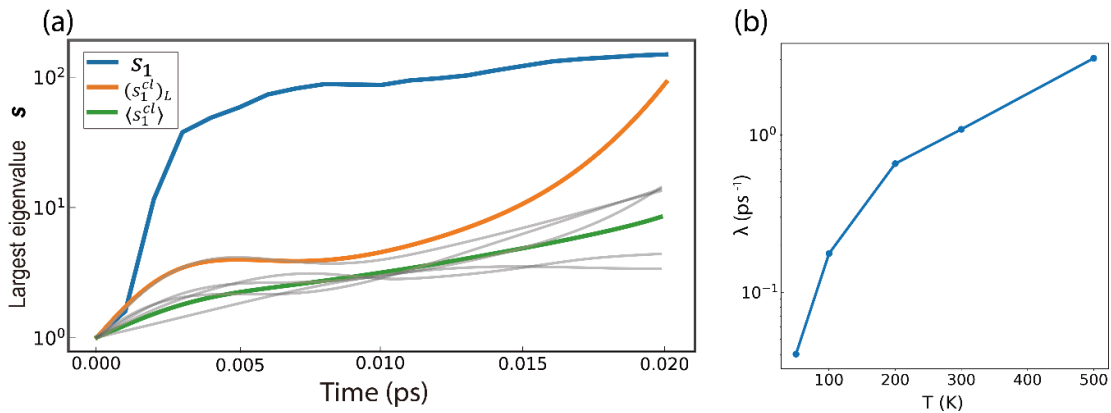

**Figure 4.** Color online. (a) The Lyapunov eigenvalues $s$ for initial state (6, 5, 1, 3, 5, 3) of thiophosgene, $SCCl_2$. Here $(s_1^{cl})_L(t)$ (orange) is the largest $s_1^{cl}(t)$ among 100 trajectories with different initial angle variables, $<s_1^{cl}(t)>$ (green) is the average of $s_1^{cl}(t)$ over 100 trajectories with random angle variables. 5 typical $s_1^{cl}(t)$ used for computing the average are shown in gray. $s_l(t)$ (blue) is the quantum mechanical result. Deviation between quantum and classical OTOCs is due to delayed delocalization of classical trajectories. Fig. S2 shows the same calculation with Coriolis coupling (constants for $SCCl_2$ are taken from Ref. [35]), which does not play a large role up to room temperature. (b) regularized thermal Lyapunov exponents $\lambda(T)$ for $SCCl_2$ molecule for $T$ in the range from 50 K to 500 K.

Fig. 4(b) shows the thermal OTOC for $SCCl_2$, indicating scrambling on times of order 0.5 to 25 ps. This range of timescales is comparable to barrier crossing times for thermal chemical reactions at temperature ranging from those of reactions in interstellar clouds at 50 K [55] to typical laboratory reactions at 200 °C; specifically for $SCCl_2$, stimulated emission pumping above the predissociation limit near 20,000 cm$^{-1}$ shows that dissociating states have lifetimes > 8 ps [56], and would be sensitive to the incomplete quantum scrambling in Figure 4(b). Even when the quantum scrambling of vibrational modes of photochemically excited molecules is very fast (see Fig. 6 in Appendix), it generally remains incomplete due to the existence of nearly conserved quantum numbers (so-called polyads) [28,57–59].

We also tested the effect of rotation-vibration coupling on $SCCl_2$ dynamics. Thiophosgene has moments of inertia and Coriolis coupling coefficients typical of small- to medium-sized organic molecules [35]. Up to $J=40$, which lies above the most probable angular momentum quantum number of $SCCl_2$ at room temperature ($J_{mp} \approx 32$), the effect of Coriolis coupling is negligible on the time scale of the vibrational anharmonicity. (See Fig. 7 in Appendix.)

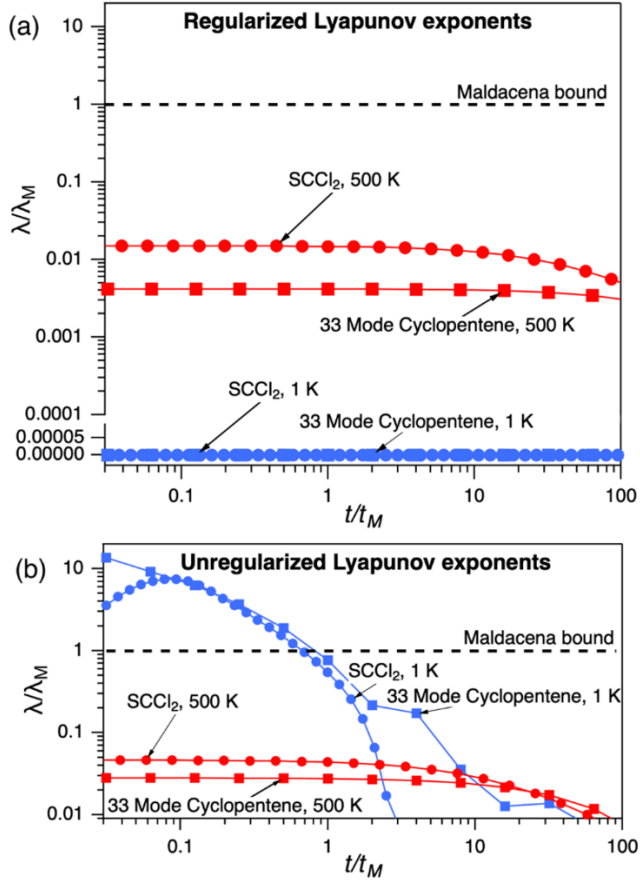

**Figure 5.** (a) Color online. Regularized thermal Lyapunov exponents scaled by the Maldacena bound for the 'all modes' models of $SCCl_2$ and 33 mode cyclopentene. The exponents obey the Maldacena bound over the whole temperature range. (b) Unregularized thermal Lyapunov exponents scaled by the Maldacena bound for $SCCl_2$ and 33 mode cyclopentene. These exponents do not obey the Maldacena bound at low temperature, as discussed in the text. Lyapunov exponents $\lambda$ are scaled by $\lambda_M = 2\pi k_B T/\hbar$, and time is scaled by $1/\lambda_M$.

### B. Thermal Quantum OTOC and compare with Maldacena bound

To compare the rate of scrambling in molecules directly with the Maldacena bound $2\pi k_B T/\hbar$ [2], we carried out thermal averaging by using the density matrix-based FEAST algorithm [60], as well as by exploiting quantum typicality [10,61,62] to reduce the thermal average to a simple expectation value (see Appendix F for details). Here we compare the thermal Lyapunov spectra derived from the unregularized OTOC $\hat{L}^{\{unreg\}}(t) = Tr\left[-[x_i(t), p_j]^2 e^{-\beta H}\right]$ to those from the regularized OTOC $\hat{L}^{\{reg\}}(t) = Tr\left[-[x_i(t), p_j] e^{-\beta H/2} [x_i(t), p_j] e^{-\beta H/2}\right]$. Results for 'all modes' $SCCl_2$ and cyclopentene are shown in Fig. 5.

The two OTOCs show quite distinct behavior at the low temperature, perhaps due to finite size effects. The Lyapunov exponents computed from $s_1(t)_{reg}$ obey the Maldacena bound at all temperatures we studied, whereas those computed from $s_1(t)_{unreg}$ do not. This point is most clearly manifested when one evaluates $\langle g | -[x_i(t), p_j]^2 | g \rangle$ for the ground state $|g\rangle$. In that case $s_1(t)_{reg}$ shows no growth, reflecting the quiescence of the dynamics near absolute zero, whereas $s_1(t)_{unreg}$ has unphysical exponential growth.

At higher temperature, where the Logan-Wolynes facile flow criterion $\rho_{loc}V_{anh} > 1$ is satisfied, the Lyapunov exponent of the unregularized OTOC approaches those of the regularized one, in harmony with what has been seen in the SYK model [10]. This can be seen in Figure 5 for the Lyapunov exponents at 500 K and in Figure 8 in Appendix F.

The question of which thermal Lyapunov spectrum gives a better measure of scrambling has been discussed in the literature [63,64], where the regularized OTOC has been argued to be more appropriate in field theory because it enforces the appropriate hermiticity properties of the operator and can be brought into correspondence with the Boltzmann many-particle transport equation. It appears that for molecules, the regularized thermal OTOC at low temperature does agree better with Maldacena's bound and with Herzfeld's notion, based on old quantum theory, of a maximum rearrangement rate. Several of the molecules we examined approach, in the number of degrees of freedom, those used in SYK simulations. Thus, molecules can be experimentally testable benchmarks for the applicability of OTOCs to many-body quantum systems of moderate size.

## IV. CONCLUSION

Our numerical explorations reveal both similarities and differences between the classical and quantum Lyapunov spectra for molecules in the size range from $f=6$ to $f=36$ degrees of freedom. Clearly one should exercise caution when assuming scrambling in molecules due to purely vibrational couplings. The three time scales manifested in the OTOC give an indication of when scrambling is good enough so that reactions may be treated statistically. For reactions faster than the ballistic time, statistical models of reaction rates cannot be employed. Only for sufficiently slow chemical reactions is a statistical treatment advised. When reactions occur during the Lyapunov and scrambling regimes, statistical theories that can take into account energy flow rates can be employed. [12,21,65,66]

The quantum OTOC levels off and deviates from its classical counterpart because of quantum interference [67], ultimately reducing the number of participating states $N_p$ below the value corresponding to the full density of states. This type of deviation has been rationalized using the semiclassical phase-space formulation [6,68], in which operators are translated into their phase-space counterparts using Wigner transforms and the Moyal expansion of the equations of motion [6]. On the other hand, both for stronger coupling (Figure 2) or when the molecule's classical phase space contains only small chaotic islands (Figure 4(a)), the classical and quantum OTOCs can differ substantially. The quantum system can undergo localization due to the finite value of $\hbar$ even when the classical system is chaotic (Figure 2) or conversely, the quantum system can sample chaotic regions early on while classical trajectories still remain trapped near invariant tori (Figure 4(a)). These effects are visible in the small isolated quantum systems considered here, but are less apparent in macroscopic systems due to rapid quantum decoherence [69].

All these examples highlight that one must be cautious in assuming that 'classical simulation is good enough for molecules with enough degrees of freedom at room temperature.' Our computational results suggest that even at chemical energy and with dozens of vibrational degrees of freedom, many molecules scramble information about their initial state slowly. Molecules could also provide a practical pathway for testing different implementations of the OTOC experimentally: The measurement of molecular OTOCs may be facilitated using entangled photons, [70,71]. The forward-backward control loop proposed by Rabitz [72] along with the quantized Ulam control conjecture [73] then open up the possibility for quantum control while the OTOC remains sufficiently small [74].

*Acknowledgements.-* C. Z. and M. G. were supported by the James R. Eiszner Chair. Computations were supported by XSEDE grant MCB180022 on Bridges/Pylon to M. G. and C. Z, funded by National Science Foundation grant ACI-1548562. P. G. W. was supported by the Center for Theoretical Biological Physics, sponsored by NSF grant PHY- 2019745. Additionally, we wish to recognize the D.R. Bullard Welch Chair at Rice University, Grant C-0016 (to Wolynes).

## Appendix A : Equations of Motion for Quantum and Classical Dynamics

We adopt the scaling Hamiltonian in [24], $H = H_0 + V$, where

$$H_0 = \sum_{i=1}^{N}(n_i + 1/2)\hbar\omega_i = \sum_i \epsilon_i(n_i) \qquad (4)$$

and

$$V = V_0 \sum_{m} \prod_{m_i} (-1)^{m_i} \left[\gamma (f_i/f_0)^{1/2}\right]^{m_i} \left(a_i + a_i^\dagger\right)^{m_i} \quad (5)$$

Here the quantum number vector is defined as $\boldsymbol{m} = \{m_1, m_2, \cdots\}$. $H_0$ describes the uncoupled motion of a set of (harmonic) oscillators with mode energies $\epsilon_i$, while $V$ describes the anharmonic couplings between the oscillators. The mode frequencies are $f_i$, with $f_0$ being median mode frequency in our model. $\gamma$ is the anharmonic scaling factor in our model. In Fock space, the index $\boldsymbol{m}$ gives the quantum number differences $m_i^+$ and $m_i^-$ between anharmonically coupled states (e.g. for two states |73> and |65>, $m_1^+$=0, $m_1^-$=1, $m_2^+$=2, and $m_2^-$=0). The sum of all differences $m_i^+$ and $m_1^-$ gives the total order of the coupling $m$. $m \geq 3$, except when the sum is 1 (corresponds to an $m = 3$ cubic coupling) or 2 (corresponds to $m = 4$ quartic coupling). The two states $|73>$ and $|65>$ in the example are coupled by an $m = 3$ cubic coupling.

We can write the above Hamiltonian in terms of $(J_i, \theta_i)$ in classical phase space:

$$H^{cl} = H_0^{cl} + V^{cl} \quad H_0^{cl} = \sum_{i=1}^{N} \omega_i (J_i + 1/2), \quad (6)$$

where

$$V^{cl} = V_0 \sum_{m} \prod_{m_i} (-1)^{m_i} \left[\gamma (f_i/f_0)^{1/2}\right]^{m_i} \left(2\sqrt{J_i/\hbar} \cos \theta_i\right)^{m_i}. \quad (7)$$

We then simulate the classical equations of motion for $H^{cl}$ using the Bulirsch-Stoer algorithm, and the time-dependent Schrödinger equation in state space for $H$ using the SUR algorithm.

## Appendix B : Quantum Lyapunov spectrum

We compute the Lyapunov spectrum as described in [36]. For our system, we reformulate the $\widehat{M}_{ij}$ in terms of ladder operators $\hat{a}$, $\hat{a}^\dagger$ associated with coordinates and momentum {x,p} using the relation[1]:

$$\hat{x} = \sqrt{\frac{\hbar}{2m\omega}} (\hat{a} + \hat{a}^\dagger) \quad (8a)$$

---

[1] We find that choosing vibrational action-angle coordinate $(J_i, \theta_i)$ with approximately conserved actions as conjugate variables to compute the OTOC is not as enlightening for molecules: the classical OTOCs $L_{ij}^{cl}$ grows as $t^2$ at early time and exponentially later, but the quantum OTOC $\hat{L}_{ij}(t)$ only shows $t^2$ growth before it levels off to a small number of participating states $N_p$ for moderate-sized molecules at the scrambling time, and no exponential growth can be observed. Defining $\widehat{M}_{ij}$ in terms of the ladder operators does produce rapid growth of the quantum operator as well.

$$i\hat{p} = \sqrt{\frac{m\omega\hbar}{2}}(\hat{a} - \hat{a}^\dagger) \tag{8b}$$

Here $\hat{M}_{ij}(t) = \frac{1}{i\hbar}[\hat{z}_i(t), \hat{z}_j]$, $z_i(i = 1, \cdots, 2N) = \{x, p\}$ can be proved equivalent to the following expression after an unitary transformation:

$$\hat{M}(t) = \begin{pmatrix} [\hat{a}_i(t), \hat{a}_j] & \cdots & [\hat{a}_i^\dagger(t), \hat{a}_j] \\ \cdots & \cdots & \cdots \\ [\hat{a}_i(t), \hat{a}_j^\dagger] & \cdots & [\hat{a}_i^\dagger(t), \hat{a}_j^\dagger] \end{pmatrix} \tag{9}$$

We denote the Lyapunov spectral matrix $\hat{L}_{ij}$ evaluated with state $|\{n\}\rangle$ as $\hat{L}_{ij}^{\{n\}}(t)$. We evaluate $\hat{L}_{ij}^{\{n\}}(t)$ by inserting a set of states $|\{n'\}\rangle$ between $\hat{M}_{ij}(t)$:

$$\hat{L}_{ij}^n(t) = \langle\{n\}|\sum_k \hat{M}_{ki}^*(t)\hat{M}_{kj}(t)|\{n\}\rangle = \sum_{\{n'\}}(\sum_k \langle\{n\}|\hat{M}_{ki}^*(t)|\{n'\}\rangle \langle\{n'\}|\hat{M}_{kj}(t)|\{n\}\rangle). \tag{10}$$

$\langle\{n'\}|\hat{M}_{kj}(t)|\{n\}\rangle$ are computed by solving time-dependent Schrödinger equation for state $|\{n\}\rangle$ and $|\{n'\}\rangle$:

$$\langle\{n'\}|\hat{M}_{kj}(t)|\{n\}\rangle = \langle\{n'\}|[\hat{a}_k(t), \hat{a}_j]|\{n\}\rangle$$
$$= (\langle\{n'\}|e^{iHt})\hat{a}_k(e^{-iHt}\hat{a}_j|\{n\}\rangle) - (\langle\{n'\}|\hat{a}_j e^{iHt})\widehat{a_k}(e^{-iHt}|\{n\}\rangle) \tag{11}$$

Owing to the local nature of the anharmonic couplings, we can limit this solution to the exploration of the Fock space starting from initial states $\boldsymbol{n} = \{n_1, n_2, \cdots, n_N\}$ to nearby states $\boldsymbol{n'} = \{n'_1, n'_2, \cdots, n'_N\}$ where the difference $|\mathbf{n} - \mathbf{n'}|$ is limited in size. We impose a 1-norm distance cutoff in state space $\|n\|_1 = \sum_{i=1}^N |n_i - n_i^{(1)}| \leq R$ to construct the computational basis set of states $\{\boldsymbol{n^{(1)}}\}$ as a local basis set. A similar cutoff $\|n'\|_1 = \sum_{i=1}^N |n_i - n'_i| \leq R'$ ($R' < R$) is imposed for choosing initial states $\{\boldsymbol{n'}\}$ (in eq. (10)) which we use to compute the average $\hat{L}_{ij}(t)$. We find that choosing $R'$, $R = 4$ to 5 is sufficiently large to ensure the convergence of the $\hat{L}_{ij}(t)$.

The time-dependent Schrödinger equation is solved using the shifted-update-rotation (SUR) algorithm [38]. The SUR algorithm belongs to the family of symplectic propagators [75] that explicitly show the correspondence of classical and quantum time evolution. We also confirmed the accuracy of $\hat{L}_{ij}(t)$ computed in this way by propagating initial states using the Chebyshev propagator [76]. The OTOCs computed using these two methods are in excellent agreement with each other.

## Appendix C: Classical Lyapunov spectrum

To compute the classical Lyapunov spectral matrix $L_{ij}^{cl}(t)$, we use the Bulirsch-Stoer algorithm to integrate the classical Hamiltonian equations of motion. For classical simulations, this algorithm for time propagation is less prone to phase errors than encountered using symplectic propagators, for which numerical errors in propagation can lead to an overestimate of the amount of scrambling. Convergence was monitored as a function of integration step size. As mentioned in the main text, for each molecular example system, we calculated classical trajectories starting from 100 initial conditions starting with random angles ($\theta_i$), but with action value ($J_i$) that correspond to the quantum numbers $\{\boldsymbol{n}\}$ of the initial state used to compute the quantum OTOC.

## Appendix D: Models

### I. Schofield Wyatt Wolynes (SWW) model

The SWW Hamiltonian is shown in [40]. It consists of 6 anharmonic oscillators with similar frequencies ~ 1000 cm$^{-1}$ coupled by a cubic coupling. For the present computation, instead of working with explicitly anharmonic oscillators as did SSW, we set the self-anharmonicity to 0 and work with anharmonically coupled pure harmonic oscillators. See Table II of [40] for parameter values. The weakly chaotic regime corresponds to a cubic anharmonic coupling strength $\phi^{(3)} = 7$.

### II. Cyclopentanone-based model

The simplified model contains the 8 vibrational modes of molecule cyclopentanone in the 2900 cm$^{-1}$ region of CH stretches and HCH bending overtones [77]: $f = (2210, 2222, 2966, 2945, 2130, 2126, 2880, 2880)\ cm^{-1}$. We use the scaling Hamiltonian based on eqs. (4) and (5) in ref. [22].

$$V_{ii'} \approx \prod_k R_k^{n_k} \tag{12a}$$

$$R_k \approx \frac{3050^{1/Q}}{270} \omega_k^{1/2} \overline{v_k^{1/2}} \tag{12b}$$

This Hamiltonian has scaling factors $R_k$ that scale with vibrational frequency $\omega_k$ and mean occupation number $v_k$. A least-squares fit to directly computed sample potential surfaces of the molecules in [24] is in good agreement with the numerical relation above. For reference, a typical third order coupling strength is $\phi_3 = 25\ cm^{-1}$ and scaling factor $\gamma \approx 0.2$.

### III. Cyclopentene and cyclopentanone full-dimensional vibrational models

All 33 vibrational frequencies with $a_1, a_2, b_1, b_2$ symmetry for cyclopentene are shown in TABLE II in [46]. The 36 vibrational frequencies with $A, B$ symmetry for cyclopentanone are shown in Table 4(a) in [77]. We use the scaling Hamiltonian which is eq. (4)(5) in [22] and we include cubic and quartic anharmonic couplings. When constructing the anharmonic coupling, we make use of the symmetry of the molecules' vibrational motion. For example, cubic coupling allows symmetry combination $(a_1, b_1, b_1)$ but $(a_1, b_1, b_2)$ is not allowed. For these relatively large organic molecules, we only managed to compute the Lyapunov spectrum for low-lying energy states, and we had to restrict the range of the 1-norm distance cutoff to $R' = 1$ instead of the value $R' = 4$ used in all other simulations.

For cyclopentene we focused on the state: (1 0 0 0 0 1 0 0 0 0 0 1 1 1 0 1 0 1 0 1 0 1 0 0 0 0 0 0 0 0 1 1 1) with energy $E = 17569 cm^{-1}$.

For cyclopentanone we focused on the state: (1 0 0 0 0 1 0 0 0 0 0 1 1 1 0 1 0 1 0 1 0 1 0 0 0 0 0 0 0 0 1 1 1 0 0 0) with energy $E = 17357 cm^{-1}$.

### IV. SCCl$_2$ molecule

We employed the Hamiltonian for SCCl$_2$ that was fitted from the spectroscopy data in ref. [35]. We find that most regions $(J_i \leq 5)$ show regular dynamics in classical simulation; however, we observed exponential growth of the Lyapunov spectrum in our quantum simulation. We singled out state (6,5,1,3,5,3) with an energy of 15000 $cm^{-1}$ to compute the quantum and classical Lyapunov spectra. For the classical Lyapunov spectrum, we used 100 points with random initial angles for the average. The points that exhibit the largest classical Lyapunov exponent lie at resonance junctions and can be found by using the Lyapunov weighted sampling technique [78,79].

## Appendix E: Lyapunov weighted sampling techniques

We followed the method described in [78,79]: instead of computing the full Lyapunov spectrum, which is expensive, we use the fast Lyapunov indicator (FLI) $\lambda^{FLI}$ [29,80] as the indicator for chaos. The specific procedure is as follows: we construct an initial ensemble with the same action $\vec{J}$ and different angles $\vec{\theta}$ and compute $\lambda^{FLI}$ for each point. Then we performed a biased random walk obeying constraint (constant $\vec{J}$) using MCMC (Monte Carlo Markov Chain) techniques with $-\lambda^{FLI}$ as the effective energy. By choosing the appropriate temperature $T$ and a reasonable step size, at the final time the ensemble will converge to a set of energy minima, which corresponds to the maximally chaotic region.

# Appendix F: Thermal OTOC

## I. Exact Diagonalization method - FEAST

The thermal OTOC is obtained by first solving for the eigenstates of the molecular Hamiltonian $H$ in an energy band using the FEAST algorithm [60] implemented in the oneAPI Math Kernel Library. Two features of FEAST make it most suitable for computing the thermal OTOC here. First, it is able to solve for eigenstates within a given energy range to save computational cost. Second, FEAST allows one to distribute tasks across parallel processors and reduce the computational burden per single process.

The regularized ($\hat{L}^{\{r\}}(t)$) and the unregularized Lyapunov spectrum ($\hat{L}^{\{u\}}(t)$) are computed as follows:

$$\hat{L}_{ij}^{\{r\}}(t) = \text{Tr}\left[\sum_k e^{-\beta H/2}\hat{M}_{ki}^*(t)e^{-\beta H/2}\hat{M}_{kj}(t)\right] = \qquad (13)$$

$$\sum_k \sum_\phi \sum_\psi e^{-\beta(E_\phi + E_\psi)/2} \langle\phi|\hat{M}_{ki}^*(t)|\psi\rangle\langle\psi|\hat{M}_{kj}(t)|\phi\rangle$$

and

$$\hat{L}_{ij}^{\{u\}}(t) = \text{Tr}\sum_k e^{-\beta H}\hat{M}_{ki}^*(t)\hat{M}_{kj}(t)] = \qquad (14)$$

$$\sum_k \sum_\phi \sum_\psi e^{-\beta E_\phi} \langle\phi|\hat{M}_{ki}^*(t)|\psi\rangle\langle\psi|\hat{M}_{kj}(t)|\phi\rangle$$

Here $|\psi\rangle, |\phi\rangle$ are eigenstates.

## II. Approximating thermal average using Haar random state

As mentioned in the main text, the thermal average can be approximated by taking the expectation value with respect to Haar-random initial states. [10] We generate the Haar-random states by drawing each (complex) element in $|\psi\rangle$ from a Gaussian distribution. Errors introduced by this approximation can be reduced by averaging over many initial Haar states. Here the final results are obtained by averaging over 5 random Haar states. In all of these setups, we can compute the thermal OTOC by solving time-dependent Schrödinger equation using the SUR or Chebyshev algorithms. Both algorithms in our computation give comparable performance when computing the thermal Lyapunov spectrum.

The regularized ($\hat{L}^{\{r\}}(t)$) and the unregularized Lyapunov spectrum ($\hat{L}^{\{u\}}(t)$) are computed as follows:

$$\hat{L}_{ij}^{\{r\}}(t) = Tr\left[\sum_k e^{-\beta H/2}\hat{M}_{ki}^*(t)e^{-\beta H/2}\hat{M}_{kj}(t)\right] \approx \qquad (15)$$

$$\sum_m \langle\psi|e^{-\beta H/4}\hat{M}_{ki}^*(t)e^{-\beta H/4}|m\rangle\langle m|e^{-\beta H/4}\hat{M}_{kj}(t)e^{-\beta H/4}|\psi\rangle$$

$$\hat{L}_{ij}^{\{u\}}(t) = \text{Tr}\left[\sum_k e^{-\beta H}\hat{M}_{ki}^*(t)\hat{M}_{kj}(t)\right] \approx \qquad (16)$$

$$\sum_m \langle\psi|e^{-\beta H/2}\hat{M}_{ki}^*(t)|m\rangle\langle m|\hat{M}_{kj}(t)e^{-\beta H/2}|\psi\rangle$$

Here $e^{-\beta H/4}|\psi\rangle$ is computed by propagating $|\psi\rangle$ in imaginary time [81].

**Table 1.** Dilution factor σ and the Logan-Wolynes facile flow criterion. The Logan-Wolynes facile flow criterion $\rho_{loc}V_{anh} > 1$ can be found in (eq.4.12 (a-c)) in ref. [82]. The dilution factor $\sigma = lim_{t\to\infty} P(t)$ is the long-time limit of survival probability: $P(t) = |\langle\psi(0)|\psi(t)\rangle|^2$.

|  | SWW (13000 $cm^{-1}$) | 8 mode cyclopentanone (9500 $cm^{-1}$) | 8 mode cyclopentanone (43700 $cm^{-1}$) | Cyclopentene (17569 $cm^{-1}$) | Cyclopentanone (17357 $cm^{-1}$) | SCCl$_2$ (15000 $cm^{-1}$) |
|---|---|---|---|---|---|---|
| σ | 0.058 | 0.1 | 0.03 | 0.03 | 0.005 | 0.004 |
| $\rho_{loc}V_{anh}$ | 3.1 | 2.5 | 4.2 | 1.84 | 10.5 | 5.5 |

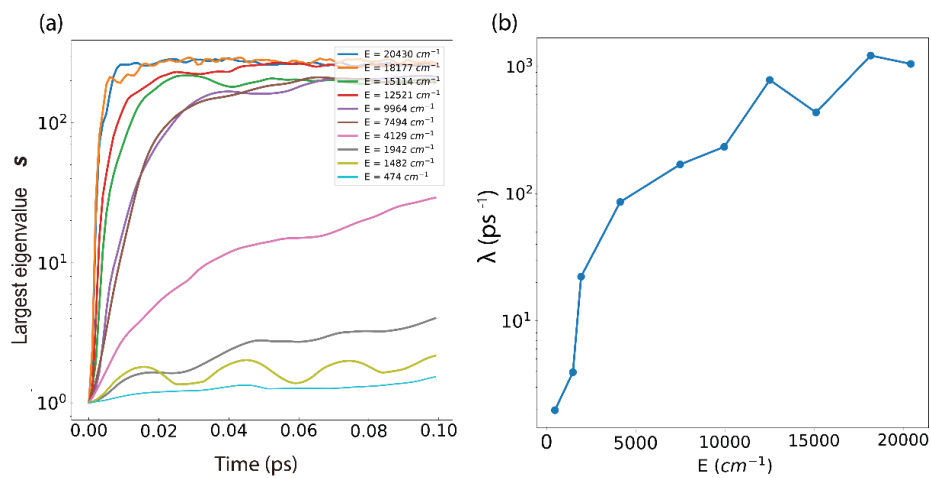

**Figure 6**. Microcanonical OTOC for SCCl$_2$. Color online. (a) OTOC for SCCl$_2$ molecules for states with energy $E \in [500, 20000]\ cm^{-1}$. (b) Lyapunov exponent λ for SCCl$_2$ molecules with OTOC shown in (a).

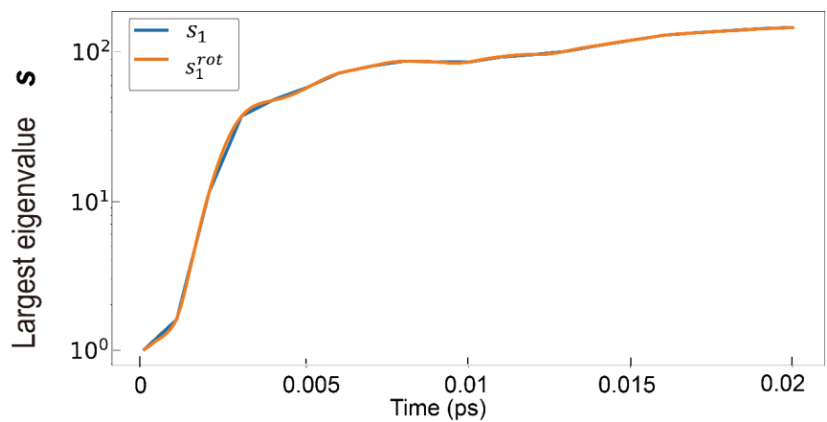

**Figure 7.** Effect of Coriolis couplings on scrambling in SCCl$_2$. Largest Lyapunov eigenvalue *s* with and without Coriolis coupling in SCCl$_2$ is shown. The blue curve is for *J*=0 in absence of Coriolis couplings, as shown in the main text. The orange curve corresponds to *J*=40, M=0, near the maximum rotational population at room temperature. The Coriolis Hamiltonian from ref. [35] was used for the calculations. The Coriolis effect mixes states on a ~100 times longer time scale than vibrational anharmonicity.

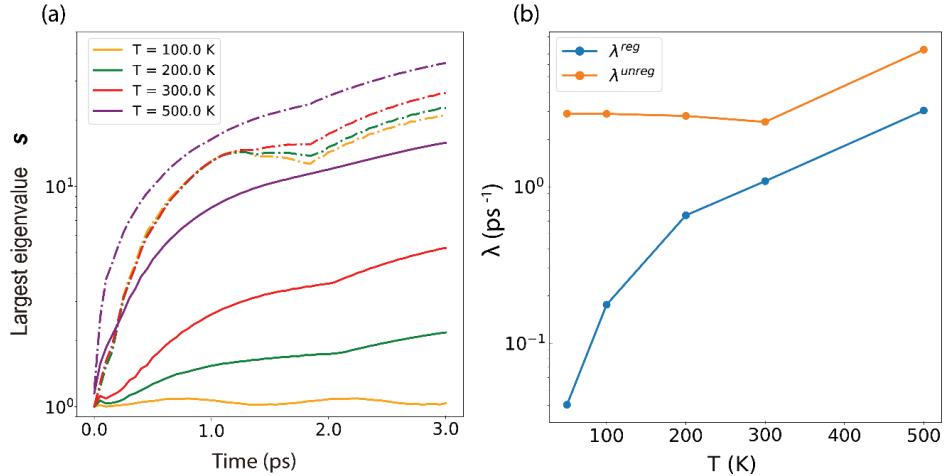

**Figure 8.** Color online. SCCl$_2$ thermal OTOC and Lyapunov exponent. **(a)** regularized (solid) and unregularized (dashdot) thermal Lyapunov eigenvalues *s* for SCCl$_2$ molecules. The regularized thermal *s* increase from the expected flat line at low T (100 K) to growing at the high *T*. In contrast, unregularized thermal *s* is much larger than its regularized counterpart at low T but these are almost identical to each other at high T. **(b)** regularized (blue) and unregularized (orange) thermal Lyapunov exponent $\lambda(T)$ for SCCl$_2$ molecules. As mentioned in the main text, at low temperature, $\lambda^{reg}$ is much smaller than $\lambda^{unreg}$. However, as *T* increases to 500 K, which is around mean vibrational frequency $\overline{\nu}$ of SCCl$_2$ molecule, $\lambda^{reg}$ and $\lambda^{unreg}$ become close to each other.